\documentstyle[epsfig]{mn}
\title[Linking Cluster Formation to Large Scale Structure]
      {Linking Cluster Formation to Large Scale Structure}
\author[J.M.\ Colberg et al.]
  {J.\ M.\ Colberg,$^1$\thanks{Email: {\tt jgc@mpa-garching.mpg.de}}
   S.\ D.\ M.\ White,$^1$ A.\ Jenkins,$^2$ F.\ R.\ Pearce$^2$\\
   $^1$Max--Planck--Institut f\"ur Astrophysik,
       Karl--Schwarzschild--Str.\ 1, D--85740 Garching\\
   $^2$Physics Dept., University of Durham, GB--Durham DH1 3LE}
\begin{document}
\newcommand{\etal}{et al.}
\maketitle
\begin{abstract}

We use two high resolution CDM 
simulations to show that (i) when clusters of galaxies form the infall 
pattern of matter is not random but shows clear features which are correlated
in time; (ii) in addition, the infall patterns are correlated with the 
cluster's surrounding Large Scale Structure; (iii) Large Scale
Structure shows a mix of both filaments and sheets;
(iv) the amount of mass in filaments is slightly 
larger for a low $\Omega$ model.

\end{abstract}
\section{Introduction}

Clusters of galaxies are well--studied despite
their rarity because they are easy to find and are the
largest known objects in approximate dynamical equilibrium. 
Furthermore, their current dynamics and their evolution
appear to be dominated by the gravity of the unseen dark 
component.  

The assumption that the formation of a galaxy cluster is dominated by 
gravity allows one to study clusters with collisionless 
N--body simulations. The dynamics is then well understood, and it
is possible to simulate different cosmologies to investigate 
how cluster formation is affected by cosmological parameters. 
Only recently has it become possible to run simulations
with high enough resolution to study details of cluster formation with
high accuracy. Currently, two techniques are used. The first runs
a cosmological N--body simulation with poor resolution, extracts a cluster from it
and re--simulates the cluster with a multi--mass technique 
which gives high resolution on
the cluster itself (for recent results see the comprehensive paper by
Huss \etal{} 1997). Simulations like these can be done on 
high--end workstations. Their main disadvantage is that each simulation 
yields only one cluster. The second approach is somewhat simpler and runs
a cosmological N--body simulation with high resolution 
throughout a large volume. 
This is only really practical on the latest generation of parallel
supercomputers. In this
case, however, one gets a large number of clusters in each simulation.

Simulations of the second type are particularly useful if one
wants to investigate how the formation of a cluster is influenced
by Large Scale Structure (LSS). This is the topic of this paper. We
will study links between the formation of clusters within large high 
resolution N--body simulations and their surrounding LSS. In addition,
we will address several aspects of the LSS itself, e.g., we will show that in
these Cold Dark Matter (CDM) universes, 
both filaments and sheets can be found, and that it is 
possible to display these directly despite the large difference in
their density. This is interesting
because of the ongoing debate about how best to describe LSS.
Sheets or walls (Geller \& Huchra 1989, de Lapparent \etal{} 1986),
filaments (Giovanelli \etal{} 1986), and mixes of these to produce
a cell--like geometry (J\^oeveer \& Einasto 1987)
have all been suggested. Bond et al.\ (1995) have shown that 
this mix already shows up in the overdensity pattern of the 
initial density field.

The outline of this paper is as follows. In section \ref{simulations}
we will introduce the simulations we use. The formation of 
clusters will be investigated in section \ref{formation}. After
selecting the clusters (sec.\ \ref{selection}), their formation
history will be constructed (sec.\ \ref{history}). This will then be
studied in detail (sec.\ \ref{investigate}) using Aitoff projection
maps and correlation functions. We link the formation process 
to LSS in section \ref{connect}. Finally, mass fractions
of the matter in the filaments are computed (sec.\ \ref{mass}).
Section \ref{summary} summarizes our results.

\section{The Simulations} \label{simulations}

Our simulations were carried out using {\sevensize HYDRA}, a parallel 
adaptive particle--particle
particle--mesh (AP$^3$M) code, developed within the VIRGO
supercomputing consortium (For details on the code c.f.\ 
Couchman \etal{} 1995 and Pearce \& Couchman 1997).
The simulations were started on the CRAY
T3D at the Computer Center of the Max Planck Society in Garching
using 128 processors. Once the clustering strength
required an even 
larger amount of total memory, they were finished on the T3D at the 
Edinburgh Parallel Computing Centre using 256 processors.

A set of four 256$^3$ particle simulations with 
different cosmological parameters was
run. These were designed primarily for another purpose and will be
described in more detail in Kauffmann \etal{} 1997.
From these two are taken, namely an $\Omega=1$ model, $\tau$CDM, with
$h=0.5$, $\Gamma=0.21$ and a
box size of 85\,$h^{-1}$\,Mpc, and an
open $\Omega=0.3$ one, OCDM with $h=0.7$, $\Gamma=0.21$ and a box size
of 141\,$h^{-1}$\,Mpc (As usual, the Hubble constant is expressed as
$H_0 = 100 h$\,km\,sec$^{-1}$\,Mpc$^{-1}$, and $\Gamma$ is the shape
parameter of the CDM spectrum (c.f.\ Bond \etal{} 1984)). The models 
are normalized such that they yield the observed {\em abundance\/} of
rich clusters by mass, i.e., $\sigma_8=0.6$ and $\sigma_8=0.85$ for $\tau$CDM 
and OCDM, respectively. All the models were run with the same phases 
so that objects are directly comparable.

These simulations are particularly useful for the purposes of this paper
because they resolve clusters and their surrounding LSS very well.
From the 50 output times which were stored we use
the ones later than $z=1$.
They are $z=$0.93, 0.82, 0.72, 0.62, 0.52, 0.43, 0.35, 0.27, 0.20,
0.13, 0.06, and 0.0.

\begin{figure*}
\begin{center}
\epsfig{file=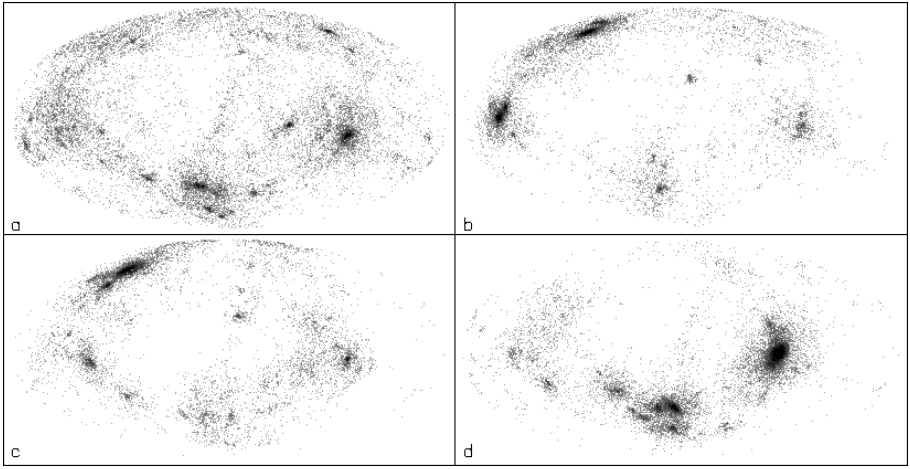,width=120mm}
\end{center}
\caption{The infall pattern of matter onto a cluster in the $\tau$CDM
model viewed from an hypothetical observer in the barycenter of
the system. Shown are different time intervals: (a) $z=0.$ to $z=0.13$,
(b) $z=0.13$ to $z=0.27$, (c) $z=0.27$ to $z=0.43$, and 
(d) $z=0.43$ to $z=0.62$.}
\label{2dfallin}
\end{figure*}

\section{The Formation of Clusters} \label{formation}

\subsection{The Selection of Clusters} \label{selection}

The definition of a cluster adopted here is analogous to
Abell's criteria for the selection of clusters from galaxy
catalogs. Of course, in a simulation clusters can be found using full
3D information. At $z=0$, high density regions are found
using a standard friends--of--friends group finder with
a linking length of 0.05 times the mean interparticle separation.
Spheres with a radius of
$r_{\rm A}\equiv1.5\,h^{-1}$\,Mpc (comoving) are put around the barycenters of 
those regions and all particles inside them are taken
as members of the cluster. If two cluster centers are closer than $r_A$
the more massive one is taken and the other one deleted from the list.
No recentering is done after such a step. From both models the ten most 
massive clusters are taken. The mass range of these clusters is from
2.7$\times 10^{14}h^{-1}M_{\odot}$ (3.5$\times 10^{14}h^{-1}M_{\odot}$) to
7.3$\times 10^{14}h^{-1}M_{\odot}$ (8.4$\times 10^{14}h^{-1}M_{\odot}$) 
for the $\tau$CDM (OCDM) model.

\subsection{Construction of the Formation History} \label{history}

After having found the clusters the particles in each of them are 
marked in a list.
Using this information they are extracted from the whole set
of particles at all redshifts. After that two additional pieces of
information are obtained for each particle: The time when it falls into
the cluster and its position at that time. This is done in the
following manner.
With the above selection criteria a cluster is a spherical object with 
a radius $r_{\rm A}$ at any time. Let the current 
time be $z_i$. Going back to $z_{i-1}$, 
some of the particles which will be inside $r_{\rm A}$ at $z_i$ are still outside.
Hence, these particles will fall into 
the cluster between $z_{i-1}$ and $z_i$. So for these particles $z_{i-1}$
as well as the position at this time are saved.
As the center of the cluster we take the barycenter at
$z_{i-1}$ of the biggest lump.
 
\subsection{Investigating the Formation of the Clusters} \label{investigate}

Often the formation of a cluster is modeled by the spherical collapse
of some high peak in the initial density field. However, previous studies,
e.g.\ Tormen \etal{} \shortcite{Tormen97}, have already shown that
the actual formation process in hierarchical models is rather
irregular. Instead of a steady accretion of matter, lumps fall onto a
pre--existing object, a process well known in and typical of CDM models.

Here we study the build--up of clusters in more detail. 
A hypothetical observer is placed at the barycenter of 
each cluster. This observer watches the matter as it falls into the cluster. 
For spherical infall he would see matter coming in more or less 
randomly from all directions.
For each observer, maps of the infall pattern are produced by plotting
the positions of the particles at infall on an
Aitoff projection. Figure \ref{2dfallin} shows typical examples of such
maps for a cluster in the $\tau$CDM simulation.
From this some points can be made. (i) It is 
obvious that matter is not falling in uniformly over the sky
as one would expect in a spherical 
infall model. Rather infall occurs from distinct directions.
(ii) There is a tight correlation between
the infall directions at different times. That is, the cluster
forms by accretion from a few preferred directions.

In order to quantify this process we have computed the 
autocorrelation function of the infalling matter
\begin{equation}
1 + \omega(\vartheta) = \frac{4\pi}{A(\vartheta)}
                        \frac{N_{\rm PP}(\vartheta)}{N(N-1)}\,,
\end{equation}
where $N_{\rm PP}(\vartheta)$ denotes the
number of particle pairs separated by an angle $\vartheta + 
\delta\vartheta$, $A(\vartheta)=2\pi \sin\vartheta {\rm d}
\vartheta$ is the size of the
annulus, and $N$ is the total number of particles in the sample. Obviously, 
$\omega(\vartheta)$ is the excess probability of finding a particle pair 
with separation $\vartheta + \delta\vartheta$
in the simulations compared with spherical infall.

\begin{figure}
\begin{center}
\epsfig{file=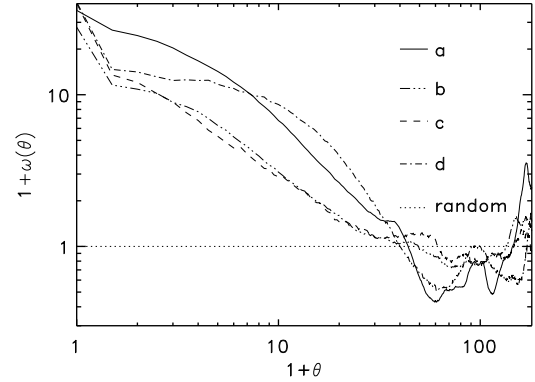,width=80mm}
\end{center}
\caption{The auto correlation functions of the matter shown in
fig.\ \ref{2dfallin} for the same time intervals used in the four
panels of fig.\ \ref{2dfallin}.
Also shown is the prediction for spherical infall.}
\label{2dfallinangcorr}
\end{figure}

Figure \ref{2dfallinangcorr} shows the autocorrelation functions
for the infalling matter of the cluster in fig.\ \ref{2dfallin} and
for random spherical infall. 
For small angles all the curves have a peak. 
This just reflects the particle clumps seen in fig.\ \ref{2dfallin}.
The strength of the peak directly reflects the amount of 
matter in these clumps. For some of the curves, 
peaks also appear at larger angles. For example, curve (b) has a second
peak around 60$^{\circ}$. This reflects the angle between the two most
massive objects in fig.\ \ref{2dfallin}(b). 

These curves can be compared directly to the 
correlations of matter between
different maps, as quantified by the cross--correlation function
\begin{equation}
1 + \tilde{\omega}(\vartheta) = \frac{4\pi}{A(\vartheta)} 
                                \frac{N_{\rm P_1P_2}(\vartheta)}{N_1\cdot N_2},
\label{ccf}
\end{equation}
where $N_{\rm P_1P_2}$ denotes the number of pairs of a
particle from map 1 and one from map 2 separated by an angle $\vartheta +
\delta\vartheta$, where $A(\vartheta)=2\pi \sin\vartheta 
{\rm d}\vartheta$ is the size
of the annulus again, and $N_1$ and $N_2$ are the number of particles in
the maps 1 and 2, respectively. 

\begin{figure}
\begin{center}
\epsfig{file=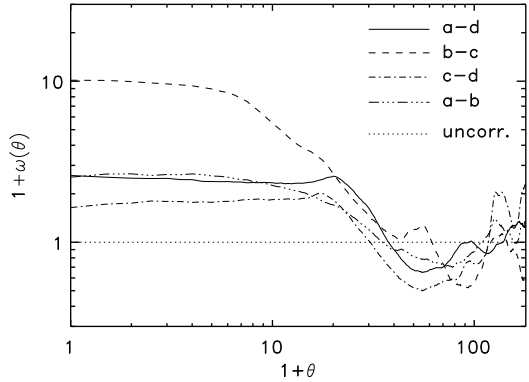,width=80mm}
\end{center}
\caption{The cross correlation between pairs of maps from
fig.\ \ref{2dfallin}. 
Also shown is the result expected for uncorrelated maps (dotted).}
\label{2dfallincrosscorr}
\end{figure}

Figure \ref{2dfallincrosscorr} shows cross--correlations between 
pairs of maps from fig.\ \ref{2dfallin}.
These have similar scale but are generally weaker than the 
autocorrelations. This can be seen
by comparing the maps directly, too. The behaviour for this
particular cluster is 
typical for both the auto-- and cross--correlations in the infall patterns of
{\em all\/} clusters in {\em both\/} the $\tau$CDM and OCDM model.
Not a single case was found which deviates qualitatively from it. 

From the above, it is apparent that correlations 
between the infall patterns at different times are strong. 
The question now is whether one should expect to
find such a correlation? From previous studies, e.g. Tormen \etal{} 1997,
it is clear that clusters form by the accretion
of haloes. This is already reflected by the discussion above. The
question left is: Why is the infall pattern of
matter between so many and so different redshift intervals
correlated? This may be rephrased as: Is there a connection between
the infall pattern and LSS itself? This will be discussed in the
next sections.

\begin{figure*}
\begin{center}
\epsfig{file=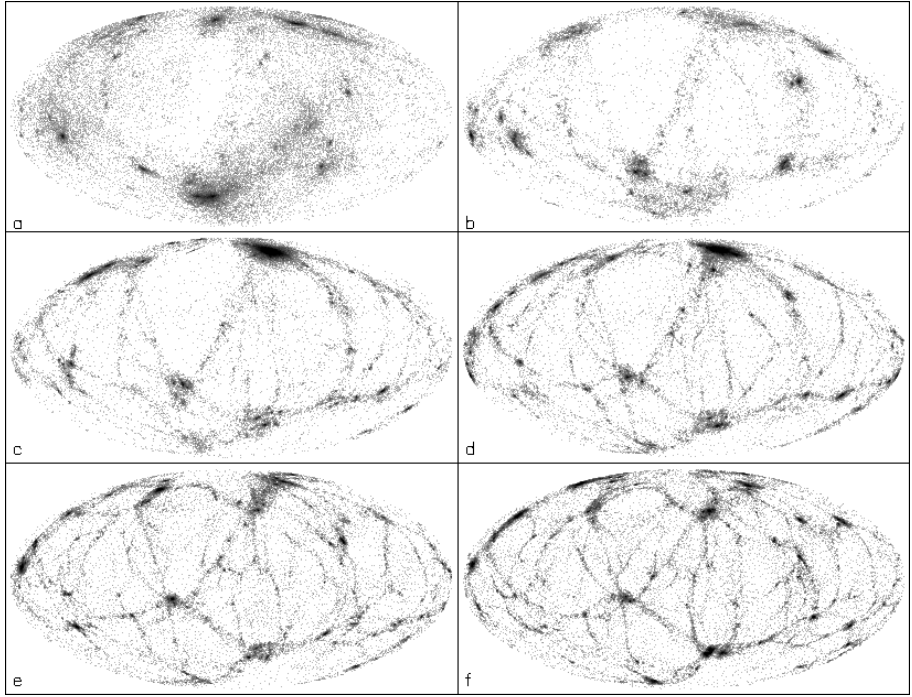,width=120mm}
\end{center}
\caption{The LSS around the cluster shown in fig.\ \ref{2dfallin}
viewed from an hypothetical observer at the barycenter of
the cluster at $z=0$. Shown are shells at
1.5 to 3.0 (a), 3.0 to 4.5 (b), 4.5 to 6.0 (c), 6.0 to 7.5 (d), 
7.5 to 9.0 (e), and 9.0 to 10.5\,$h^{-1}$\,Mpc (f) from the cluster center.}
\label{2dLSS1}
\end{figure*}

\begin{figure*}
\begin{center}
\epsfig{file=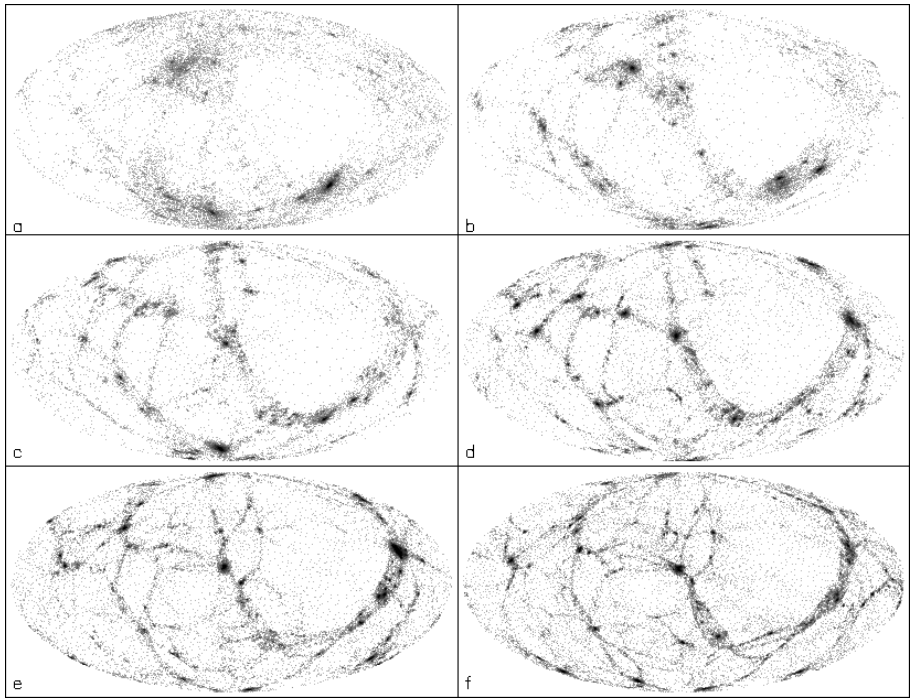,width=120mm}
\end{center}
\caption{The LSS around another cluster from the $\tau$CDM
simulation shown using the same representation as in fig.\ \ref{2dLSS1}.}
\label{2dLSS2}
\end{figure*}

\subsection{Connecting Cluster Formation and Large Scale Structure} \label{connect}

From the above it is obvious that the during formation of a cluster 
matter falls in from well--defined directions. 
What is the connection between these
dierections and Large Scale Structure? In order to investigate 
this the distribution of matter surrounding the clusters at $z=0$ is obtained as
follows. Around the clusters we put onion--like shells of thickness 
$1.5\,h^{-1}$\,Mpc are put. All particles in a shell are extracted. The
hypothetical observer at the cluster center draws maps of these
particles, i.e., LSS is viewed from the center of each cluster.

\begin{figure}
\begin{center}
\epsfig{file=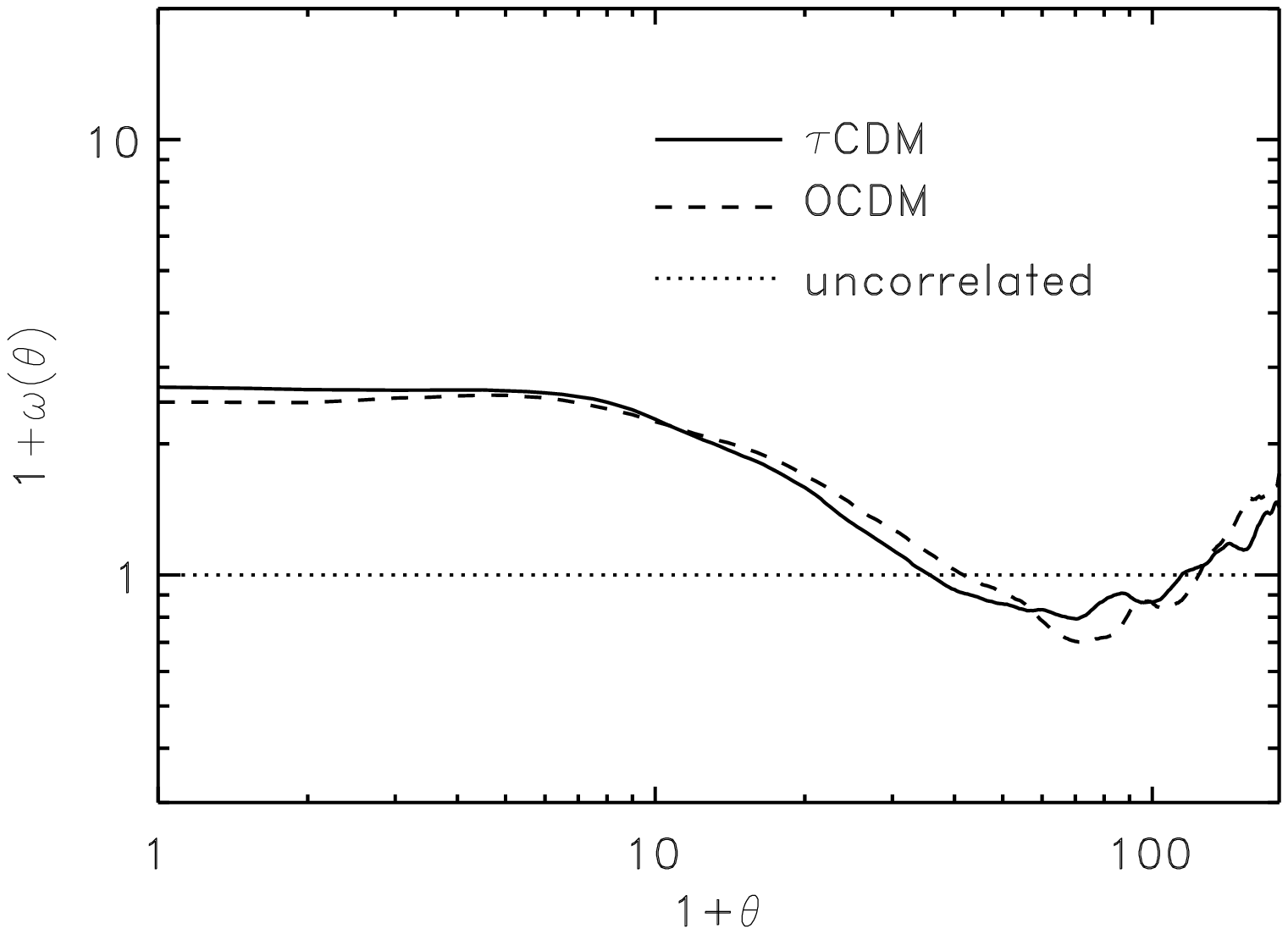,width=80mm}
\end{center}
\caption{The combined cross correlation between the infall patterns of
the clusters in the $\tau$CDM and OCDM simulations and their surrounding LSS.}
\label{Crosscorr}
\end{figure}

Figure \ref{2dLSS1} shows maps for 
shells surrounding the $\tau$CDM cluster which we 
analyzed in figures \ref{2dfallin} to \ref{2dfallincrosscorr}. Again, these
maps are typical of those found for all the clusters.
From the maps some points can be made. (i)
There exist density enhancements in the distribution of the particles
which only marginally change their locations from map to map. 
Most of them are
more or less circular. These must clearly be filaments extending
outwards from the cluster. In addition, fig.\ \ref{2dLSS1} shows another
interesting feature. There are enhancements which connect the filaments and 
also extend outwards from map to map, 
but are less dense. Figure \ref{2dLSS2} shows the LSS around
a different $\tau$CDM cluster where these connections between the filaments are
very strong. There is a U--shaped broad band in the right part of
five of the six maps. This object is obviously a sheet.
One has to note that filaments can be
found in {\em all\/} of the cluster maps. There are connections
between them in all maps, too. However, impressive examples of sheets like the
one in figure \ref{2dLSS2} are rare. (ii) When comparing the maps in figure
\ref{2dLSS1} with the one in fig.\ \ref{2dfallin}
one can see that the big clumps fall in along the filaments.
This is not only true for the lowest redshift range but for the earlier ones,
too. Even at a redshift of 0.6 infall onto clusters is
tightly coupled to LSS at $z=0$. 

In order to
quantify this connection a little bit further, the angular cross correlation
functions between the combined infall maps of the cluster and the LSS 
maps are computed for each cluster in each cosmology. 

Figure \ref{Crosscorr} shows cross correlations between the
infall patterns and surrounding LSS for the ten $\tau$CDM and OCDM 
clusters. These are averaged over the redshift ranges and radii shown 
in fig.s \ref{2dfallin}, \ref{2dLSS1}, and \ref{2dLSS2}.
Very similar cross correlations are found for other time and radius
intervals. This mean cross correlation
behaves in a similar manner for the cross correlations
between the different maps in fig.\ \ref{2dfallincrosscorr}. There
is indeed a well defined correlation between the infall
onto clusters and their surrounding LSS. This correlation does
not depend on $\Omega$.

\begin{figure*}
\begin{center}
\epsfig{file=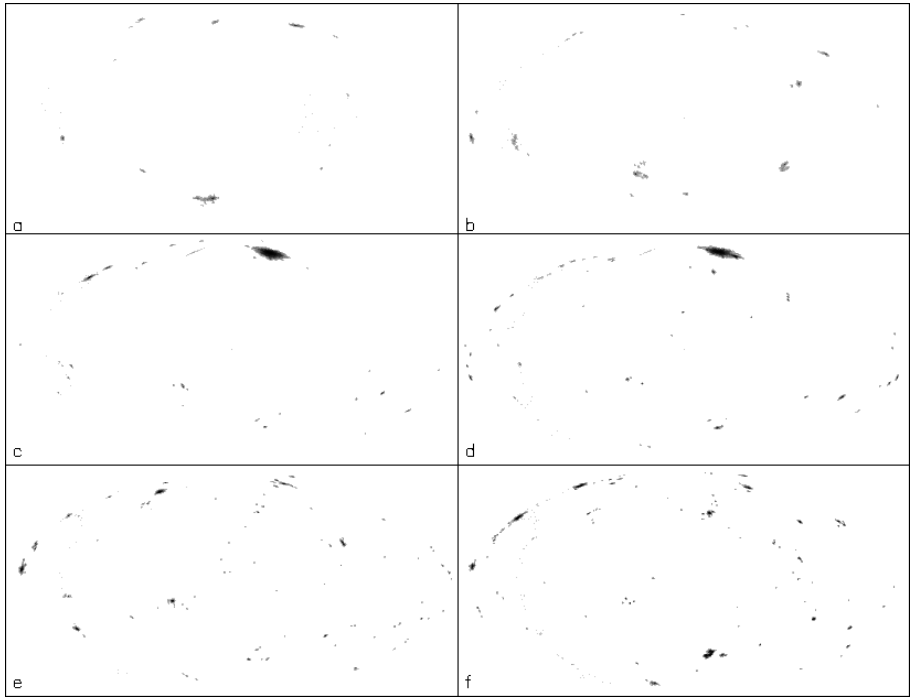,width=120mm}
\end{center}
\caption{A different representation of fig.\ \ref{2dLSS1} where only
the lumps (filaments) show up.}
\label{2dLSS1fils}
\end{figure*}

\subsection{Fraction of Mass in the Peaks} \label{mass}

What is the amount of mass which can be seen in the various 
structures in the above maps?
In order to answer this question we have computed the amount of mass 
inside the dark spots. This is done by using a standard
friends--of--friends group finder on the sets of points on the unit
sphere from which the above maps were drawn. 
As linking parameter a value of $b=0.2$ times the mean
interparticle separation is taken. All objects with 20 or more
particles are considered as big groups.

\begin{table}
\caption{Fractions of mass in the dark spots 
for the maps in fig.\ \ref{2dLSS1} and
\ref{2dLSS2}. Given are the total mass fractions ($f_{\rm all}$)
as well as the mass fractions of the two biggest spots
($f_1$ and $f_2$) for the different radii.}
\begin{center}
\begin{tabular}{lrrrrrr}
Radius & \multicolumn{3}{c}{Fig. \ref{2dLSS1}} & 
    \multicolumn{3}{c}{Fig. \ref{2dLSS2}} \\
$[$Mpc/h$]$ & $f_{\rm all}$ & $f_1$ & $f_2$ & $f_{\rm all}$ & $f_1$ &
    $f_2$ \\
1.5 -- 3.0  & 11\% & 5\%  & 2\% & 13\% & 9\%  & 2\% \\ 
3.0 -- 4.5  & 22\% & 3\%  & 3\% & 24\% & 4\%  & 4\% \\
4.5 -- 6.0  & 52\% & 42\% & 2\% & 28\% & 10\% & 3\% \\
6.0 -- 7.5  & 45\% & 32\% & 1\% & 34\% & 5\%  & 4\% \\
7.5 -- 9.0  & 32\% & 7\%  & 4\% & 43\% & 17\% & 3\% \\
9.0 -- 10.5 & 39\% & 8\%  & 5\% & 32\% & 6\%  & 2\% \\
\end{tabular}
\end{center} 
\end{table}

Table 1 gives the fraction of mass inside such big dark spots for
the figures \ref{2dLSS1} and \ref{2dLSS2} ($f_{\rm all}$). 
Also shown is the fraction of mass in the two most massive spots
in each map ($f_1$ and $f_2$). 
Typically, about a third of the mass lies in filaments at the
overdensity of $\sim$20 picked out by our choice of linking
length.

Fig.\ \ref{2dLSS1fils} shows a representation of the maps in fig.\ 
\ref{2dLSS1} where only the particles in these dark spots, i.e.,
the filaments are plotted. From this it is apparent that
filaments are clumpy structures rather than homogeneous 
cylinders.

Performing a similar analysis for the infall patterns onto clusters
gives results which vary more strongly between clusters
and time intervals. E.g., for the cluster shown
in fig.\ \ref{2dfallin} the fractions of mass in the dark spots
are 5\%, 15\%, 30\%, and 51\% for the maps (a) to (d), respectively.
This scatter between 5\% and around 55\% was quite typical for 
clusters in both the $\tau$CDM and the OCDM sample.

For the whole $\tau$CDM (OCDM) cluster sample the averaged mass 
fractions in the filaments
are 14\% (14\%) , 29\% (35\%), and 42\% (46\%) for
shells beginning at radii 1.5, 3.0, and 4.5\,$h^{-1}$\,Mpc. These
stay constant at around 40\% (48\%) for larger radii.
There is a slightly larger mass fraction in the filaments 
in the low $\Omega$ model. 

\section{Conclusions} \label{summary}

Our study of the formation process of clusters and its connection with LSS
has yielded several conclusions. (i) In CDM universes, clusters form by the
accretion of collapsed haloes onto other 
pre--existing haloes. This occurs from
preferred directions. These directions do not
change much with time. (ii) There is a correlation
between the formation process of a cluster and its surrounding LSS. 
Qualitatively speaking, matter falls in mainly from filaments 
and sheets. 
The filaments show up as clear density enhancements in our 2D projections. 
They extend outwards from the cluster center and are connected
by less dense sheets of matter. 
Because of their considerably lower density contrast these sheets are
nearly impossible to find in 3D representations of N--body simulations.
Our representation clearly shows that both filaments
and sheets do exist in simulations. Quantitatively speaking, the amount of
mass in the filaments is around 40\% and 48\% of the total mass
for radii from 4.5 to 10.5\,$h^{-1}$\,Mpc in the $\tau$CDM and OCDM
model, respectively. At smaller radii, it is around 30\%. However,
the mass distribution is dominated by lumps inside the filaments. 
(iii) The only difference we could find between the $\tau$CDM
and the OCDM model is in the amount of mass in the filaments, it is
slightly larger for the OCDM model.

The formation process of each cluster is governed by its surrounding
LSS. The internal properties of the cluster may change during its formation,
as shown by Tormen \etal{} \shortcite{Tormen97}, but this is no
completely chaotic process but 
is linked to the LSS. LSS itself appears to be a mix of both filaments and
sheets in CDM universes at least in the representation we
have emphasized in this paper. These simulation results clearly
reinforce the Cosmic Web picture of structure formation
proposed by Bond et al.\ (1995)

\section*{Acknowledgements}

JMC wishes to thank Volker Springel, Adi Nusser, Ravi Sheth, Alexander Knebe, and
Antonaldo Diaferio for many useful and encouraging discussions.

The simulations were carried out on the Cray T3D's at the
Computer Center of the Max--Planck--Gesellschaft in Garching
and at the Edinburgh Parallel Computing Centre. The analysis was done on the
IBM SP2 at the Computer Center of the Max--Planck--Gesellschaft 
in Garching.


\begin{thebibliography}{99}
%
\bibitem[\protect\citename{Bond \& Efstathiou }{1984}]{Bond84}
Bond J.R., Efstathiou G., ApJ, 285, L45 (1984)
%
\bibitem[\protect\citename{Bond }{1995}]{Bond95}
Bond J.R., Kofman L., Pogosyan D., astro-ph/9512141
%
\bibitem[\protect\citename{Couchman }{1995}]{Couchman95} 
Couchman H.M.P., Thomas P.A., Pearce F.R., ApJ, 452, 797 (1995)
%
\bibitem[\protect\citename{de Lapparant }{1986}]{deL86} 
de Lapparant V., Geller M.J., Huchra J.P., ApJ, 302, L1 (1986)
%
\bibitem[\protect\citename{Geller }{1989}]{Geller89} 
Geller M.J., Huchra J.P., Science, 246, 897 (1989)
%
\bibitem[\protect\citename{Giovanelli }{1986}]{Giovanelli86} 
Giovanelli R., Haynes M., Myers S.T., Roth J., AJ, 92, 250 (1986)
%
\bibitem[\protect\citename{Huss }{1997}]{Huss97}
Huss A., Jain B., Steinmetz M., submitted to MNRAS
%
\bibitem[\protect\citename{J\^oeveer }{1978}]{Joeveer78}
J\^oeveer M., Einasto J., in IAU Symp.\ 79, The Large--Scale
Structure of the Universe, eds.\ Longair M.S., Einasto J. 
(Dordrecht: Reidel), 241 (1978)
%
\bibitem[\protect\citename{Kauffmann }{1997}]{Kauffmann97} 
Kauffmann G.A.M., Colberg J.M., Diaferio A., White S.D.M., in preparation
%
\bibitem[\protect\citename{Pearce }{1997}]{Pearce97} 
Pearce F.R., Couchman H.M.P., Hydra: A parallel
adaptive grid code, submitted to New Astronomy (1997)
%
\bibitem[\protect\citename{Tormen }{1997}]{Tormen97} 
Tormen G., Bouchet F.R., White S.D.M., MNRAS, 286, 865 (1997)
%
\end{thebibliography}
\end{document}